\documentclass[a4paper]{elsarticle}

\usepackage{lineno,hyperref}
\modulolinenumbers[5]

\usepackage{graphicx}
\usepackage{dcolumn}
\usepackage{bm}
\usepackage{multirow}
\usepackage{textcomp}
\usepackage[]{subfig}

\usepackage{booktabs}
\usepackage{array}
\usepackage{pgf}
\usepackage{graphicx}
\usepackage[absolute,overlay]{textpos}
\usepackage{calc}
\usepackage{amsmath}
\usepackage{amsfonts}
\usepackage{amssymb}
\usepackage{textcomp}
\usepackage{tikz}
\usepackage{gensymb}
\usepackage{float}
\usepackage{amsmath}
\DeclareMathOperator{\Tr}{Tr} 

\usepackage{algorithm, algorithmic} 

\renewcommand{\algorithmiccomment}[1]{\bgroup\hfill$\triangleright$~#1\egroup}
\usepackage{color}
\definecolor{Sun}{rgb}{0.164,0.126,0.322}
\definecolor{Green}{rgb}{0,0.300,0.300}
\definecolor{Red}{rgb}{0.4,0,0}
\definecolor{Grey}{RGB}{105,105,105}
\definecolor{White}{rgb}{1,1,1}

\usepackage{soul}

\usepackage{csquotes}

\newcommand{\commentCH}[1]%
{\textsf{\textcolor{blue}{#1$^{\mathrm{CH}}$}}}
\newcommand{\commentAV}[1]%
{\textsf{\textcolor{red}{#1$^{\mathrm{AV}}$}}}
\newcommand{\correction}[1]%
{\textsf{\textcolor{red}{#1}}}

\journal{Computer Methods in Applied Mechanics and Engineering}
\begin{document}

\begin{frontmatter}

\title{A damage and failure implementation for the simulation of ductile solids with Total-Lagrangian Smooth Particle Hydrodynamics.}

\author[monash]{A.~de Vaucorbeil\corref{cor1}}
 \ead{alban.devaucorbeil@monash.edu}
\author[monash]{C. R. Hutchinson}

\cortext[cor1]{Corresponding Author}

\address[monash]{Department of Materials Science and Engineering, Monash University, Clayton 3800, VIC, Australia}

\begin{abstract}

Smooth-Particle-Hydrodynamics is gaining popularity for the simulation of solids subjected to machining, wear, and impacts. Its attractiveness is due to its abilities to simulate problems involving large deformations resulting from the absence of mesh, recent improvements in stability conferred by the development of Total-Lagrangian version of SPH (TLSPH), but also its availability in the open-source software LAMMPS. This implementation features a damage model similar to the ``pseudo-spring'' method which creates instabilities when used for the simulation of ductile materials.
In this contribution, we present a new damage and failure model for TLSPH suitable for ductile materials. In this implementation, not only the constitutive equations but also the TLSPH approximation are modified in order to take into account the change in material properties as well as the presence of discontinuities due to the initiation and growth of damage. This new approach is accompanied by the implementation of the Cockroft-Latham, Johnson-Cook, and Gurson-Tvergaard-Needleman damage criteria. The predictive capabilities of the implementation of this new damage model are then tested and compared against both experimental and results of Finite Element simulations.
\end{abstract}


\end{frontmatter}

\date{\today}

\section{Introduction}

Smooth Particle Hydrodynamics (SPH) is a meshfree particle method originally developed in the seventies for modeling three-dimensional problems in astrophysics~\cite{GM77, L77}. Recent extensions now allow SPH to be used to simulate solid mechanics problems~\cite{LESL07, VF08}. Its principle is the use of a discrete kernel function centered about particles to determine, at their location point, an approximation of continuous fields such as density, velocity, stresses, and internal energy.
This principle, when applied to solid body deformations transforms the underlying set of partial differential equations into simple algebraic equations.  

Due to its abilities to simulate problems involving large deformations, SPH is gaining popularity for the study of the deformations of solids in the fields of machining~\cite{LESL07, VF08}, wear~\cite{LVEVRG16, VLRR17,TP12}, and resistance to impacts~\cite{ICSR17, XD17, CS15, CS13}. Such popularity has also been fueled by its recent implementation in the open-source particle code LAMMPS~\cite{Pl95}. This implementation is shipped in the user package called Smooth Mach Dynamics (SMD) created by Ganzenm\"uller~\cite{G14} and is based on the numerically stable and efficient Total-Lagrangian version of SPH (TLSPH). 

In cases involving large deformations and damage, true meshfree particle methods such as SPH have several advantages over mesh-based methods such as Finite Element Method (FEM). One advantage is the absence of need to mesh, or remesh when large distortions occur. Another advantage is its the ability to realistically simulate damage and fracture due to the absence of immutable connectivity between neighbouring particles~\cite{RL96}. In fact, the decrease or loss of interaction between two neighbouring particles could be seen respectively as the presence of damage or as the formation of a crack between them. This concept is exploited in the so called ``pseudo-spring'' damage and fracture method~\cite{CS13}. In this method, the kernel function in the direction of every virtual segment \textit{linking} two neighbouring particles, called a ``pseudo-spring'', is scaled according to the amount of damage present in that ``pseudo-spring''.
When the amount of damage reaches the critical limit, the kernel function in the ``pseudo-spring's'' direction becomes null, and the connectivity between particles is lost. This method is similar to the use of cohesive elements~\cite{OP99} in mesh-based methods which are particularly adapted to the simulation of debonding, and failure of brittle solids. 

However, the ``pseudo-spring'' fracture model has three limitations. 
\begin{enumerate}
\item It imposes the calculation of a quantity in a non-integration point (\textit{i.e.} between particles). 
\item It breaks the mathematical approximation upon which SPH is built by scaling the kernel functions. 
\item It imposes particles to be linked to their immediate neighbours only, as done by Chakraborty and Shaw~\cite{CS13}, or the interaction with non-intermediate neighbours separated from that particle by a crack to be scaled appropriately~\cite{VV13}.
\end{enumerate}

The other currently available damage and fracture implementation for SPH is the particle cracking method~\cite{RB04,RB07,RZBNX10}. Contrary to the ``pseudo-spring'' method, damage is calculated at the center of particles (\textit{i.e.} the integration points). When damage reaches a critical limit, the particle is split into two. In this case, damage is not localized as its value corresponds to an average over each particle's corresponding volume. Therefore, the precise location of cracks is not defined and the way particles need to be split is non-trivial. 

 Currently, the only damage models implemented in TLSPH LAMMPS package for solids were models for brittle failure using a version of the ``pseudo-spring'' failure method~\cite{G14}. No damage law for ductile material has been fully implemented. Varga, Leroch and coworkers \cite{LVEVRG16, VLER17, VLRR17} have used the TLSPH LAMMPS package to simulate the mechanical response of a ductile material to scratch, without taking into account damage. By recalculating the particles' neighbour lists after a few timesteps, they have relied on numerical failure, more specifically the loss of connectivity between neighbouring particles as their distance increase, to account for damage. This so-called ``natural ability'' for SPH to account for damage has also been used by other researchers~\cite{LESL07, TP12, XD17, CS15, CS13}. However, this is unsatisfactory when it comes to predict the mechanical response of ductile materials experiencing damage.

No damage law for ductile material is currently available in LAMMPS. Only an undocumented pre-implementation of Johnson-Cook damage law is present but is flagged as \textit{currently unsupported}. Moreover, when used with the ``pseudo-spring'' failure method, this implementation suffers instability issues triggered by the start of the propagation of failure. 

In this paper, a new implementation for damage and failure for the LAMMPS TLSPH package is introduced. In this approach, no alteration of the kernel functions is used, and the particles' neighbour list remains identical throughout each simulation. Thus, this approach does not rely on numerical fracture to account for damage. Instead, the inherent displacement discontinuity generated by the presence of cracks or voids is taken into account by scaling the velocity difference between neighbouring damaged particles. Damage is computed at the center of particles where cracks are assumed to be located.
This new method is accompanied by the implementation of three damage laws: Cockroft-Latham~\cite{CL68}, Johnson-Cook~\cite{JC85}, and Gurson-Tvergaard-Needleman~\cite{T89}. The predictive capabilities of the new implementation are then tested and compared against both experimental and Finite Element simulations.

\section{Total-Lagrangian SPH theory for solids}

\begin{figure}
  \begin{center}
    \includegraphics[height=5cm]{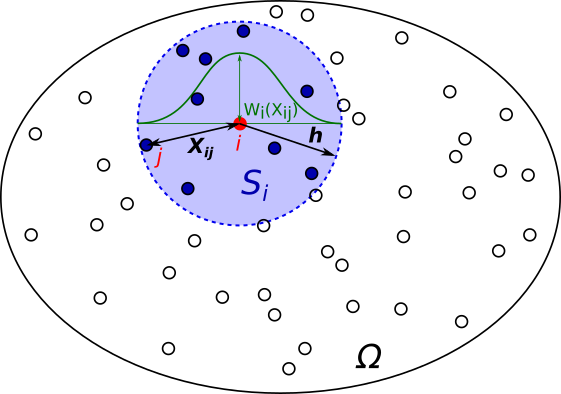}
    \caption{Schematic of total-Lagrangian SPH interactions for bulk particles in the reference configuration. A domain $\Omega$ is discretized into a set of particles with coordinate $\bm{X}$. Each particle $i$ interacts with a limited number of neighbours, denoted $j$, contained into a compact support $S_i$ centered around $i$ and limited by a radial cutoff $h$. Particle interactions are weighted by a kernel function $W_{i}(X_{ij})$ which depends on the distance $X_{ij}$ between the neighbouring particles.}
    \label{fig:SPH_interactions_bulk}
  \end{center}
\end{figure}

Smooth Particle Hydrodynamics is a particle based method in which a continuous region of solid is discretized into point particles to which a certain volume is attached. Each particle $i$ interacts with its neighbours ($S_i$) by means of a smoothing kernel function $W_i$ centered around each particle (Fig.~\ref{fig:SPH_interactions_bulk}).

In the original formulation, the kernel function travels with the particle, making it Eulerian. Unfortunately, this formulation suffers from a series of instabilities and low accuracy issues when applied to solid mechanics. The first issue SPH suffered from was the so-called ``tensile instability''~\cite{SHA95} which leads to particle clunging and numerical fracture. This was found to be caused by the use of Eulerian kernel functions~\cite{BGLX00} and was addressed by the development of Total-Lagrangian SPH (TLSPH)~\cite{BK01} in which kernel functions are fixed to the particles in their reference configuration, making the kernel functions Lagrangian. However, TLSPH was impaired by a rank-deficiency problem, which necessitates artificial particle velocity damping, and also by yet another tensile instability arising when the difference between the current and the reference configurations gets too large~\cite{BK02, RBX04, VRC06}. Rank-deficiency is solved via the use of artificial particle velocity damping, while the new tensile instability was addressed by the development of the hourglass control scheme similar to that used in Finite Elements Methods (FEM)~\cite{G15}. 

Another problem crippling the use of Eulerian kernel functions for the simulation of solids is the risk of numerical fracture under large deformations. This occurs when the distance between two neighbouring particles in the undeformed configuration becomes larger than the kernel cutoff radius and these particles are no longer neighbours in the deformed state. If total damage has not been reached, this would lead to numerical failure. The use of Lagrangian kernel functions also solves this problem. In fact, in TLSPH, the SPH approximation is always determined in the reference (undeformed) configuration. Thus, particles which are neighbours in the underformed configurations will always remain as such.

All these improvements featured in the previous implementation of TLSPH for solids in LAMMPS are detailed in the work by Leroch \emph{et al.}~\cite{LVEVRG16}.

\subsection{Governing equations}

The solid mechanics problem to solve here is described by a set of conservation equations in Lagrangian. This methodology uses a reference configuration of the simulated domain for the computation of stresses and accelerations. Alongside the unchangeable reference configuration, the current set (\textit{i.e.} deformed domain) is adjusted according to the accelerations computed in the reference configuration. Therefore, all the constitutive and conservation equations are expressed in terms of the reference coordinates $\bm{X}$, which in these simulations are taken as the initial configuration, which is supposed undeformed.

The displacement $\bm{u}$ being given by:
\begin{equation}
\bm{u} = \bm{x} - \bm{X},
\end{equation}
the respective conservation equations for mass, impulse, and energy in the total-Lagrangian framework are~\cite{G15}:
\begin{align} 
\rho J &=  \rho_0 \label{eq:conservation1}\\ 
\ddot{\bm{u}} &=  \frac{1}{\rho_0} \bm{\nabla}_0.\bm{P}^T\label{eq:conservation2}\\
\dot{e} &= \frac{1}{\rho_0} \bm{\dot{F}}: \bm{P},\label{eq:conservation3}
\end{align}
where $\rho$ is the mass density, $\bm{P}$ the first Piola-Kirchhoff stress tensor, $e$ the internal energy per unit mass, and $\bm{\nabla}$ the gradient or divergence operator. The subscript $0$ designate that the quantity or operator is evaluated in the reference configuration, and when no subscript is indicated the evaluation is in the current configuration. Finally, $J$ designates the determinant of the deformation gradient $\bm{F}$ given by:
\begin{equation}
\bm{F} = \frac{d\bm{x}}{d\bm{X}} = \frac{d\bm{u}}{d\bm{X}} + \bm{I},
\label{eq:deformation_gradient}
\end{equation}
which can be interpreted as the transformation matrix that describes the rotation and stretch of a line element from the reference configuration to the current (deformed) configuration.

\subsection{SPH approximation in the bulk in the absence of damage}
\label{subsec:building_principle}

The basis of the original SPH formulation is that the value of any continuous field function $\mathrm{f}$ at a point $\bm{x}$ can be approximated as $\langle \mathrm{f}(x) \rangle$ by the following integral interpolant~\cite{M05}: 
\begin{equation}
\langle \mathrm{f}(\bm{x})\rangle = \int_{\Omega}\mathrm{f}(\bm{x'}) W(\bm{x'} - \bm{x}, h) d\bm{x'}
\label{eq:concept}
\end{equation}
where $h$ is the cutoff radius, and the function $W$ is the kernel which is even, normalized, and has a compact support (\textit{i.e.} $W(\bm{x'} - \bm{x}, h) = 0$ for $||\bm{x'} - \bm{x}||\geq h$). Therefore, $W$ fulfills the following conditions:
\begin{equation}
\int_{\Omega} W(\bm{x'} - \bm{x}, h) d\bm{x'} = 1 \label{eq:condition_1}
\end{equation}
\begin{equation}
\int_{\Omega} (\bm{x'} - \bm{x}) W(\bm{x'} - \bm{x}, h) d\bm{x'} = 0\label{eq:condition_2}
\end{equation}

From these equations, it can be seen that the Dirac function is a valid kernel function. If used, the integral interpolant reproduces exactly the function $\mathrm{f}$. It can also be seen that if the sphere of radius $h$ centered on $\bm{x}$ is astride a boundary, the kernel function being truncated by the boundary, the conditions given by Equations \ref{eq:condition_1} and \ref{eq:condition_2} are no longer satisfied~\cite{LL10}. 

Since the kernel has a compact support, in the absence of damage, the approximation of the spatial gradient of $f$ is~\cite{LLL03Nov}:
\begin{equation}
\langle \nabla \mathrm{f}(\bm{x})\rangle = \int_{\Omega}\mathrm{f}(\bm{x'}) \nabla W(\bm{x'} - \bm{x}, h) d\bm{x'}.
\label{eq:approx_spatial_gradient_Eulerian}
\end{equation}
This can be expressed in the reference configuration and yield the Lagrangian  approximation of the spatial gradient:
\begin{equation}
\langle \nabla_0 \mathrm{f}(\bm{X})\rangle = \int_{\Omega}\mathrm{f}(\bm{X'}) \nabla W(\bm{X'} - \bm{X}, h) d\bm{X'}.
\label{eq:approx_spatial_gradient_Lagrangian}
\end{equation}

In the second step of the SPH approximation, the continuous simulated domain $\Omega$ is discretized into particles. Integrals are replaced by their discretized form written as a summation over the neighboring particles. The particle approximation of Equation \ref{eq:approx_spatial_gradient_Lagrangian} evaluated at the particle $i$ yields:
\begin{equation}
\bm{\nabla}_0\mathrm{f}(\bm{X}_i) = \sum_{j \in S_i}V_j^0\mathrm{f}(\bm{X}_j) \bm{\nabla}W_i(X_{ij}),
\label{eq:SPH_gradient_approx}
\end{equation}
where $S_i$ is the set of neighbouring particles of $i$, $X_{ij} = ||\bm{X}_j -\bm{X}_i||$, and the definition of the gradient of the kernel function is:
\begin{equation}
\bm{\nabla}W_i(X_{ij}) = \frac{dW_i(X_{ij})}{dX_{ij}}\frac{\bm{X}_j - \bm{X}_i}{X_{ij}}.
\label{eq:SPH_gradient_kernel}
\end{equation}

This formulation, however, does not fulfill zeroth- and first-order completeness conditions, \textit{i.e.} the approximation of zeroth- and first-order polynomials are not exactly approximated. This is due to the non respect, in general, of the following conditions:
\begin{align}
\sum_{j \in S_i}V_j^0W_i(X_{ij}) &= 1\\
\sum_{j \in S_i}V_j^0\bm{\nabla}W_i(X_{ij}) &= 0\label{eq:1st_order_completeness_condition}
\end{align}

In order to restore zeroth-order completeness, Monaghan~\cite{M88} introduced the following \textit{ad-hoc} improvement by adding Eq. \ref{eq:1st_order_completeness_condition} to Eq. \ref{eq:SPH_gradient_approx}:
\begin{equation}
\bm{\nabla}_0\mathrm{f}(\bm{X}_i) = \sum_{j \in S_i}V_j^0 \left( \mathrm{f}(\bm{X}_j) - \mathrm{f}(\bm{X}_i)\right) \bm{\nabla}W_i(X_{ij}).
\label{eq:0th_order_complete_SPH_gradient_approx}
\end{equation}

On the other hand, the correction for first-order completeness can be done using a corrected kernel gradient as detailed in Bonet and Lok~\cite{BL99}:
\begin{equation}
\bm{\nabla}_0\mathrm{f}(\bm{X}_i) = \sum_{j \in S_i}V_j^0 \left( \mathrm{f}(\bm{X}_j) - \mathrm{f}(\bm{X}_i)\right) \bm{\nabla}W_i(X_{ij})\bm{L}_{i}^{-1},
\label{eq:1st_order_complete_SPH_gradient_approx}
\end{equation}
where $\bm{L}_{i}$ is a correction matrix given by:
\begin{equation}
\bm{L}_{i} = \sum_{j\in S_i^i} V_j^0 (\bm{X}_j - \bm{X}_i) \otimes \bm{\nabla} W_i(X_{ij})
\label{eq:correction_matrix}
\end{equation}

\section{SPH approximation at the boundaries}

The original SPH formulation has been derived in the bulk for continuous field functions. At the boundaries, however, the field functions are discontinuous. Thus, there, the TLSPH approximation given by Eq. \ref{eq:1st_order_complete_SPH_gradient_approx} might not be valid. Derivation of an SPH approximation at the boundaries can be done following the work of Liu \emph{et al.}~\cite{LLL03Nov} who have introduced a new formulation which applies for problems with or without discontinuities called DSPH (for Discontinuous SPH).

The original 1D derivation of the DSPH approximation by Liu \emph{et al.}~\cite{LLL03Nov} is as follows. Assuming that the kernel support for the particle $i$ is bounded by $a$, and $b$ , and that the function $f$ had an \textit{integrable} discontinuity in at $d$, as shown in Fig. \ref{fig:discontinuity_1D}, the integral interpolant of $f$ can be divided into two parts:
\begin{figure}
  \begin{center}
    \includegraphics{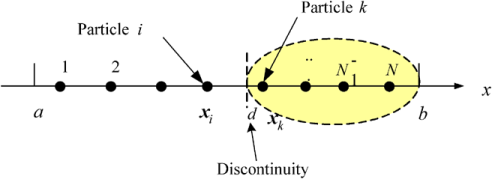}
    \caption{Particle approximations in 1D for a function with a discontinuity at point $d$. The arbitrary point $x_k$ is here associated with the nearest particle on the right hand side of the discontinuity (\textit{i.e.} particle $k$). The total number of particles in the support domain of $[a, b]$ is $N$.~\cite{LL10}}
    \label{fig:discontinuity_1D}
  \end{center}
\end{figure}
\begin{equation}
\int_a^b \mathrm{f}(x) W_i(x) dx =  \int_a^d \mathrm{f}(x) W_i(x) dx + \int_d^b \mathrm{f}(x) W_i(x) dx.
\end{equation}

Performing a second order Taylor series expansion of $f$ around $x_i$, and an arbitrary point $x_k$ located on the opposite side of the discontinuity from $x_i$ (\textit{i.e.} $d \leq x_k \leq b$) gives:
\begin{equation}
\begin{split}
\int_a^b \mathrm{f}(x) W_i(x) dx &=  \mathrm{f}(x_i)\int_a^d W_i(x) dx + \mathrm{f}(x_k)\int_d^b W_i(x) dx \\
&+ \mathrm{f}'(x_i)\int_a^d (x - x_i) W_i(x) dx \\
&+ \mathrm{f}'(x_k)\int_d^b (x - x_k) W_i(x) dx + o(h^2).
\end{split}
\end{equation}

Rearranging by combining similar terms yields:
\begin{equation}
\begin{split}
\int_a^b \mathrm{f}(x) W_i(x) dx &=  \mathrm{f}(x_i)\int_a^b W_i(x) dx + (\mathrm{f}(x_k) - \mathrm{f}(x_i))\int_d^b W_i(x) dx \\
&+ \mathrm{f}'(x_i)\int_a^b (x - x_i) W_i(x) dx \\
&+ \int_d^b [(x - x_k)\mathrm{f}'(x_k) - (x-x_i)\mathrm{f}'(x_i)] W_i(x) dx + o(h^2),
\end{split}
\label{eq:discontinuous_integral_interpolant}
\end{equation}
from which Liu \emph{et al.}~\cite{LLL03Nov} extracted $\mathrm{f}'(x_i)$ and obtained its DSPH approximation by replacing the integrals by their particle approximations.

Here, we apply the same idea but to the integral interpolant of the gradient of $\mathrm{f}$ which yields:
\begin{equation}
\begin{split}
\int_a^b \mathrm{f}(x) &\nabla W_i(x) dx =  \mathrm{f}(x_i)\int_a^b \nabla W_i(x) dx + (\mathrm{f}(x_k) - \mathrm{f}(x_i))\int_d^b \nabla W_i(x) dx \\
&+ \mathrm{f}'(x_i)\int_a^b (x - x_i) \nabla W_i(x) dx \\
&+ \int_d^b [(x - x_k)\mathrm{f}'(x_k) - (x-x_i)\mathrm{f}'(x_i)] \nabla W_i(x) dx + o(h^2),
\end{split}
\label{eq:discontinuous_integral_interpolant_gradient}
\end{equation}
and its Lagrangian particle approximation version: 
\begin{equation}
\begin{split}
\sum_{j=1}^N V_j^0 \mathrm{f}_j \nabla W_i(X_{ij}) &=  f_i\sum_{j=1}^N V_j^0 \nabla W_i(X_{ij}) + (f_k - f_i)\sum_{j=k}^N V_j^0 \nabla W_i(X_{ij})\\
&+ \mathrm{f}'_i\sum_{j=1}^N V_j^0 (X_j - X_i) \nabla W_i(X_{ij})\\
&+ \sum_{j=k}^N V_j^0 [(X_j - X_k)\mathrm{f}'_k - (X_j-X_i)\mathrm{f}'_i] \nabla W_i(X_{ij}),
\end{split}
\label{eq:particle_approximation_discontinuous_integral_interpolant}
\end{equation}
which interpolated into 3D space gives:
\begin{equation}
\begin{split}
\sum_{j=1}^N V_j^0 \mathrm{f}_j &\bm{\nabla} W_i(X_{ij}) =  \mathrm{f}_i\sum_{j=1}^N V_j^0 \bm{\nabla} W_i(X_{ij}) + (\mathrm{f}_k - \mathrm{f}_i)\sum_{j=k}^N V_j^0 \bm{\nabla} W_i(X_{ij})\\
&+ \bm{\nabla}\mathrm{f}_i\sum_{j=1}^N V_j^0 (\bm{X}_j - \bm{X}_i) \otimes \bm{\nabla} W_i(X_{ij})\\
&+ \bm{\nabla}\mathrm{f}_k \sum_{j=k}^N V_j^0 (\bm{X}_j - \bm{X}_k)\otimes\bm{\nabla} W_i(X_{ij}) - \bm{\nabla}\mathrm{f}_i\sum_{j=k}^N V_j^0 (\bm{X}_j-\bm{X}_i) \otimes \bm{\nabla} W_i(X_{ij}).
\end{split}
\end{equation}

From this equation, the kernel approximation of the gradient of $\mathrm{f}$ in $i$ is extracted and yields
\begin{equation}
\begin{split}
\bm{\nabla}\mathrm{f}_i  &= \sum_{j\in S_i} V_j^0 (\mathrm{f}_j - \mathrm{f}_i) \bm{\nabla} W_i(X_{ij})\bm{L}_{i}^{-1} - (\mathrm{f}_k - \mathrm{f}_i)\sum_{j \in S_i^k} V_j^0 \bm{\nabla} W_i(X_{ij})\bm{L}_{i}^{-1}\\
&- \bm{\nabla}\mathrm{f}_k \sum_{j \in S_i^k} V_j^0 (\bm{X}_j - \bm{X}_k)\otimes\bm{\nabla} W_i(X_{ij})\bm{L}_{i}^{-1},
\end{split}
\label{eq:DSPH_approx_gradient}
\end{equation}
where $\bm{L}_{i} = \sum_{j\in S_i^i} V_j^0 (\bm{X}_j - \bm{X}_i) \otimes \bm{\nabla} W_i(X_{ij})$, $S_i^i$ and $S_i^k$ are respectively the part of the support of $W_i$ located on the same side as and the opposite side of the discontinuity from the particle $i$.

\begin{figure}
  \begin{center}
    \includegraphics[height=5cm]{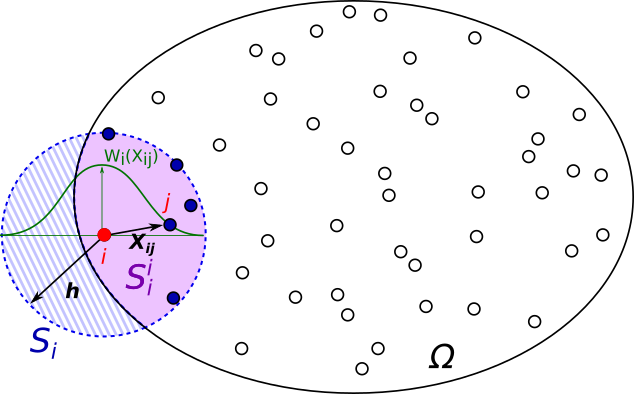}
    \caption{Schematic of Discontinuous SPH interactions at a boundary in the reference configuration. $S_i$ represents the support of the kernel function $W_i$ centered around the particle $i$. Here, $S_i$ is sitting astride a boundary. $S_i^i$ is the portion of $S_i$ located within the domain $\Omega$.}
    \label{fig:DSPH_interactions_boundaries}
  \end{center}
\end{figure}

Following the same methodology, but by splitting the interval $[d,b]$ into as many segments as there are particles, one can show that the formulation for the Lagrangian DSPH approximation of the kernel gradient is:
\begin{equation}
\bm{\nabla}\mathrm{f}_i  = \sum_{j\in S_i^i} V_j^0 (\mathrm{f}_j - \mathrm{f}_i) \bm{\nabla} W_i(X_{ij})\bm{L}_{i}^{-1}.
\label{eq:DSPH_approx_gradient2}
\end{equation}
This equation also applies in the case where no discontinuities are present and also in the case of a boundary. When no discontinuity is present $S_i^i$ is identical to $S_i$, and this formulation is identical to the Total-Lagrangian SPH approximation in Eq. \ref{eq:1st_order_complete_SPH_gradient_approx}. When a boundary is present, such as due to a crack or a void, $S_i^i$ is the portion of $S_i$ included within the domain $\Omega$ (Fig. \ref{fig:DSPH_interactions_boundaries}).

The deformation gradient and its time derivative are obtained by calculating the derivative of the displacement and velocity field, respectively (see Equation~\ref{eq:deformation_gradient}). Using the Lagrangian DSPH approximation (Equation~\ref{eq:DSPH_approx_gradient2}), they are calculated as:
\begin{align}
\bm{F}_i  &= \sum_{j\in S_i^i} V_j^0(\bm{u}_j - \bm{u}_i) \otimes \bm{\nabla} W_i(X_{ij})\bm{L}_{i}^{-1} + \bm{I} \label{eq:DSPH_approx_F}\\
\dot{\bm{F}}_i  &= \sum_{j\in S_i^i} V_j^0(\bm{v}_j - \bm{v}_i) \otimes \bm{\nabla} W_i(X_{ij})\bm{L}_{i}^{-1}, \label{eq:DSPH_approx_Fdot}
\end{align}
where $\otimes$ is the dyadic product of two vectors, and $\bm{v}$ is the particle velocity.

The nodal forces are derived from the conservation of energy following Bonnet and Lok~\cite{BL99}. Noting that the variation of the internal energy is
\begin{equation}
\dot{e} = \sum_{i\in \Omega} \frac{1}{m_i}\bm{f}_{i} \bm{v}_i,
\label{eq:var_internal_energy}
\end{equation}
and recalling Equations \ref{eq:conservation3} and \ref{eq:Crack_approx_Fdot} for the energy conservation and $\dot{\bm{F}}$, respectively, the following equality is obtained:
\begin{equation}
\dot{e} = \sum_{i\in \Omega} \sum_{j\in S_i^i} \frac{1}{m_i}V_i^0V_j^0 (\bm{v}_j - \bm{v}_i) \otimes \bm{\nabla} W_i(X_{ij})\bm{L}_{i}^{-1}:\bm{P}_i.
\end{equation}
Rearranging the summations involved and noting the anti-symmetry property of the kernel function $\bm{\nabla} W_i(X_{ij}) = - \bm{\nabla} W_j(X_{ji})$:
\begin{equation}
\dot{e}  = \sum_{i\in \Omega} \sum_{j\in S_i^i} \frac{1}{m_i}V_i^0V_j^0 \left[\bm{P}_i\bm{L}_{i}^{-T} + \bm{P}_j\bm{L}_{j}^{-T}\right]:\bm{v}_i \otimes \bm{\nabla} W_i(X_{ij})
\end{equation}
Comparing this expression with Equation \ref{eq:var_internal_energy} gives the internal forces applied onto the particle $i$ as
\begin{equation}
\bm{f}_{i} = \sum_{j\in S_i^i} V_i^0V_j^0 \left[\bm{P}_i\bm{L}_{i}^{-T} + \bm{P}_j\bm{L}_{j}^{-T}\right]\bm{\nabla} W_i(X_{ij}).
\label{eq:DSPH_nodal_forces}
\end{equation}
Owing to the anti-symmetry property of the kernel function mentioned above, this force expression will conserve linear momentum exactly, as $\bm{f}_{i} = -\bm{f}_{j}$.

\subsection{TLSPH approximation with damage}

Implementing damage into TLSPH begins by appropriately scaling the constitutive laws similarly to what is done in FEM. If the TLSPH approximation is kept unchanged, once a particle is fully damaged, it will continue applying a force on its neighbours. In fact, if $i$ is a neighbour of a fully damaged particle $j$, the force applied onto $i$ by $j$ would be:
\begin{equation}
\bm{f}_{ij} = V_i^0V_j^0 \bm{P}_i\bm{L}_{i}^{-T} \bm{\nabla} W_i(X_{ij}) \neq \bm{0},
\label{eq:fij_j_fully_damaged}
\end{equation}
similarly, the deformation matrix and its time derivative at $i$ would still be dependant of $j$.

In order to solve this problem, two methods have been used. First, the ``pseudo-spring'' method consists of the scaling of the kernel function by a function of $D_i$ and $D_j$, damages at the particle $i$ and $j$, respectively. This function is chosen such that it is equal to $1$ when both $D_i=1$ and $D_j=1$, and $0$ when either $D_i=1$ or $D_j=0$. In pure elasticity, this does not cause any problem as it is physically equivalent to a total unloading of the virtual $i$-$j$ ``link'', \textit{i.e.} the contribution of $j$ to the deformation and velocity gradient matrices at $i$ are respectively $\bm{F}_{ij} = \bm{I}$ and $\dot{\bm{F}}_{ij} = \bm{0}$. However, when plastic deformation has occurred, permanent deformation of this virtual link remains and $\bm{F}_{ij}$ cannot be equal to $\bm{I}$. Therefore, this explains why using the ``pseudo-spring'' method generates instability when used for the simulation of ductile materials.
The second method consists of splitting any fully damaged particles. This raises the difficult question as to in how many particles should the broken ones be spit into. It would also generate implementation problems within LAMMPS. This is why it was necessary to develop a new approach.

Here the chosen approach consist of scaling the difference of velocity between $i$ and $j$, \textit{i.e.} $v_j - v_i$, by a factor $1-D_j$. This is equivalent to having a crack located at the center of the particle $j$, and having the part of the volume corresponding to $j$ adjacent to $i$ moving at the same velocity as $i$. This is indeed the case in the limit of particles with infinitely low mass.

The amount of damage $D_j$ represents here the average damage in the volume $V_j$ \textit{attached} to the particle $j$. $D_j$ is evaluated at its center point and varies between 0 (undamaged) and 1 (fully damaged). When a particle is fully damaged, it has lost all stress carrying capacities but is not deleted. For practical reasons only, its velocity is set to $0$. When $j$ is undamaged, its contribution to the deformation matrix and its time derivative are respectively $\bm{F}_{ij} = V_j^0(\bm{u}_j - \bm{u}_i)\bm{\nabla} W_i(X_{ij})\bm{L}_{i}^{-1} + \bm{I}$  and $\dot{\bm{F}}_{ij} = V_j^0(\bm{v}_j - \bm{v}_i)\bm{\nabla} W_i(X_{ij})\bm{L}_{i}^{-1}$. When $j$ is fully damaged, they become respectively $\bm{F}_{ij} = \textrm{const.}$  and $\dot{\bm{F}}_{ij} = 0$. Assuming that $\dot{\bm{F}}_{ij}$ is linearly depending on $D_j$ yields:
\begin{equation}
\dot{\bm{F}}_i  = \sum_{j\in S_i^i} V_j^0 (1 - D_j) (\bm{v}_j - \bm{v}_i) \otimes \bm{\nabla} W_i(X_{ij})\bm{L}_{i}^{-1},
\label{eq:Crack_approx_Fdot}
\end{equation}
and after time integration:
\begin{equation}
\bm{F}_i  = \sum_{j\in S_i^i} V_j^0 \int_0^t(1 - D_j) (\bm{v}_j - \bm{v}_i)dt \otimes \bm{\nabla} W_i(X_{ij})\bm{L}_{i}^{-1} + \bm{I}
\label{eq:Crack_approx_F}
\end{equation}

Using Equation \ref{eq:Crack_approx_Fdot} and recalling the energy conservation equations, the nodal forces when particles are damaged are expressed as
\begin{equation}
\bm{f}_{i} = \sum_{j\in S_i^i} V_i^0V_j^0 \left[(1-D_j)\bm{P}_i\bm{L}_{i}^{-T} + (1-D_i)\bm{P}_j\bm{L}_{j}^{-T}\right]\bm{\nabla} W_i(X_{ij}).
\end{equation}

One can see that if $D_j=1$, $\bm{f}_{ij}=0$ since $\bm{P}_j = \bm{0}$, and that similarly to Equation \ref{eq:DSPH_nodal_forces}, this expression conserves linear momentum exactly.




\section{Constitutive model}

In order to close the set of Equations \ref{eq:conservation1} to \ref{eq:conservation3}, we modified the constitutive relationships used by Leroch \emph{et al.}~\cite{LVEVRG16} to take into account the effects of damage. 
The symmetric stress tensor is expressed as the sum of its isotropic part, \textit{i.e.} the hydrostatic pressure ($\sigma_m$), and the traceless symmetric deviatoric stress $\bm{\sigma_d}$:
\begin{equation}
\bm{\sigma} = \sigma_m\bm{I} + \bm{\sigma_d}.
\end{equation}

The hydrostatic pressure, on the one hand, is estimated using an equation of state (EOS). Multiple EOS are implemented in the SMD package (see the SMD documentation for more details~\cite{G14}), but as Leroch \emph{et al.}~\cite{LVEVRG16}, we use here the Mie-Gr\"uneisen equation:
\begin{equation}
\begin{cases}
\sigma_m &= \frac{\rho_0(1-D){c_0}^2(\eta -1)[\eta - \frac{\Gamma_0}{2}(\eta - 1)]}{[\eta - S_{\alpha}(\eta-1)]^2} + \Gamma_0e; \quad \eta = \frac{\rho(1-D)}{\rho_0}\textrm{ if } \sigma_m>0\\
\sigma_m &= \frac{\rho_0{c_0}^2(\eta -1)[\eta - \frac{\Gamma_0}{2}(\eta - 1)]}{[\eta - S_{\alpha}(\eta-1)]^2} + \Gamma_0e; \quad \eta = \frac{\rho}{\rho_0}\textrm{ otherwise}
\label{eq:pH}
\end{cases}
\end{equation}
where $c_0$ is the bulk speed of sound, $\Gamma_0$ the  Gr\"uneisen Gamma in the reference state. 

The deviatoric stress, on the other hand is determined using empirical models for the von Mises flow stress. Two elasto-plastic models were already implemented~\cite{G14}, namely the linear plastic and the Johnson-Cook models. Additionally, the Swift and Voce models were added to this list. Here, however, similarly to Leroch \emph{et al.}~\cite{LVEVRG16}, we use the empirical Johnson-Cook model for the von Mises flow stress scaled with damage\cite{JC85}:
\begin{equation}
\sigma_f(\varepsilon_p, \dot{\varepsilon}_p, T) = \left[A+B\left(\varepsilon_p\right)^n\right]\left[1+Cln\dot{\varepsilon}_p^*\right]\left[1 - {T^*}^m\right] (1-D),
\label{eq:JC_viscoplastic_model}
\end{equation}
where $\varepsilon_p$ is the equivalent plastic strain which is calculated jointly with the deviatoric stress tensor according to the algorithm developed by Leroch \emph{et al.}~\cite{LVEVRG16} unless the Gurson-Tvergaard-Needleman (GTN) model is used (see Algorithm \ref{alg1}), $\varepsilon_p^*$ is the normalized plastic strain rate, $T^*$ the normalized temperature, $A$ the yield stress, $B$ and $n$ the strain hardening parameters, $C$ the strain rate parameters, and $m$ the temperature coefficient. The normalized plastic strain rate and temperature are respectively given by:
\begin{align}
\dot{\varepsilon}_p^* &= \dot{\varepsilon}_p / \dot{\varepsilon}_0\\
T^* &= (T - T_0)/(T_m - T_0),
\end{align}
where $\dot{\varepsilon}_p$ and $\dot{\varepsilon}_0$ are the plastic strain rate, and the user-defined reference plastic strain rate, $T_0$ is the reference temperature, and $T_m$ the melting temperature.

\begin{algorithm}[H] 
\small
\caption{Plasticity algorithm proposed by Leroch \emph{et al.}\cite{LVEVRG16}.} 
\label{alg1} 
\begin{algorithmic}[1] 
  \STATE $\bm{\sigma}_{trial}^d = \bm{\sigma}_{n}^d + 2G(1-D)\Delta\bm{\varepsilon}^d$\COMMENT{purely elastic stress deviator update}
  \STATE $\sigma_{trial}^{eq} = \sqrt{\frac{3}{2} \bm{\sigma}_{trial}^d:\bm{\sigma}_{trial}^d}$\COMMENT{equivalent von Mises trial stress}
  \IF[yielding did not occur, purely elastic step]{$\sigma_{trial}^{eq} < \sigma_f$}
      \STATE $\bm{\sigma}_{n+1}^d = \bm{\sigma}_{trial}^d$ \COMMENT{keep trial deviatoric stress}
  \ELSE[yielding has occurred]
      \STATE $\Delta \varepsilon_{p} = \frac{\sigma_{trial}^{eq} - \sigma_{y}}{3G}$ \COMMENT{compute the plastic strain increment}
      \STATE $\varepsilon_{n+1}^p = \varepsilon_{n}^p + \Delta \varepsilon_{p}$\COMMENT{update the undamaged matrix plastic strain}
      \STATE $\bm{\sigma}_{n+1}^d = \frac{\sigma_{f}}{\sigma_{trial}^{eq}} \bm{\sigma}_{trial}^d$ \COMMENT{scale deviatoric stress back to yield surface}
  \ENDIF
\end{algorithmic}
\end{algorithm}

\section{Damage and failure models}

In order to determine the amount of damage in each particle, we have implemented in the LAMMPS SMD package\cite{G14} three different phenomenological models: Cockcroft-Latham (CL) for its simplicity~\cite{CL68}, Johnson-Cook (JC)~\cite{JC85} and Gurson-Tvergaard-Needleman (GTN) for their wide application to damage and ductile solids~\cite{T89}. CL and JC are models that only predict the damage onset point and need to be coupled with a damage evolution model, while the GTN model describes the nucleation, growth of voids until fracture occurs.
Here two variables related to damage are used: the damage initiation variable $D_{init}$, and the damage variable $D$.

\subsection{Cockcroft-Latham}
Cockcroft-Latham~\cite{CL68} proposed that the onset of fracture be simply governed by the amount of ``plastic work'' per unit volume expressed as:
\begin{equation}
W_{p} = \int_0^{\varepsilon_{eq}} \langle\sigma_1\rangle d\varepsilon_{eq},
\label{eq:cockcroft-latham}
\end{equation}
where $\sigma_1$ is the maximum principal stress, and $\langle\sigma_1\rangle = \sigma_1$ if $\sigma_1 > 0$ and $\langle\sigma_1\rangle = 0$ otherwise. Failure occurs when $W_p$ reaches a critical value $W_{cr}$ which means that the damage initiation variable is simply derived as:
\begin{equation}
D_{init} = \frac{W_p}{W_{cr}},
\end{equation}
and indicates, when equal to 1, when damage starts.

\subsection{Johnson-Cook}
The Johnson-Cook failure model~\cite{JC85} is a widely used model by both scientists and in industry to simulate the damage of ductile materials. It is available in most of the commercial and open-source finite element solvers. It is a strain rate and temperature dependant phenomenological model based on the local accumulation of plastic strain. According to this model, damage starts when~\cite{DBHL06}:
\begin{equation}
D_{init} = \sum\frac{\Delta \varepsilon_{eq}}{\varepsilon_{f}} = 1,
\end{equation}
with
\begin{equation}
\varepsilon_{f} = \left(D_1 + D_2\exp(D_3\sigma^{*})\right)(1+\dot{\varepsilon}_{p}^*)^{D_4}(1+D_5T^*),
\end{equation}
and where $D_1$, \dots, $D_5$ are five material constants, $\sigma^*$ is the stress triaxiality. 

\subsection{Damage evolution model}
Cockcroft-Latham and Johnson-Cook models describe only the point at which damage initiates. In order to have a complete model of the fracture phenomenon, a damage evolution model is required.

Here, for the sake of simplicity, it was arbitrarily assumed that the damage variable is given by:
\begin{equation}
D = 
\begin{cases}
0 &\textrm{ when } 0 \leq D_{init} < 1\\
10 \left(D_{init} - 1\right)&\textrm{ when } D_{init} \geq 1\\
\end{cases}
\end{equation}

Even if this assumption affects the details of the damage propagation, it does not change the fundamentals of the implementation on which we focus here.

\subsection{Gurson-Tvergaard-Needleman}

The Gurson-Tvergaard-Needleman (GTN) model is one of the most referenced void-nucleation-growth-coalescence model. First, Gurson~\cite{G77} developed a yield function for porous materials. Tvergaard and Needleman then extended it to fully quantitatively describe void nucleation and fracture~\cite{TN84}. This model supposes that the yield function $\Phi$ of a porous material with a void fraction $f$ is expressed as
\begin{equation}
\Phi = \left(\frac{\sigma_{eq}}{\sigma_M}\right)^2 + 2 q_1 f^* \cosh \left( \frac{3}{2} q_2 \sigma^*\frac{\sigma_{eq}}{\sigma_M} \right) - (1+q_1^2 {f^*}^2),
\label{eq:yield_GTN}
\end{equation}
where $\sigma_{eq}$ the equivalent von Mises stress, $\sigma_M = \sigma_f(\varepsilon_M^p, \dot{\varepsilon}_p, T)$ is the flow stress of the undamaged matrix (the part of material free of void) calculated according to Equation \ref{eq:JC_viscoplastic_model} , and $f^*$ the void fraction function defined as\cite{T89}:
\begin{equation}
f^* = \begin{cases}
f &\quad \textrm{for } f\leq f_c\\
f_c + \frac{1/q_1  - f_c}{f_f - f_c} (f-f_c) &\quad \textrm{for } f_c\leq f \leq f_f,
\end{cases}
\label{eq:void_fraction_function}
\end{equation}
with $f_c$ the critical void fraction, and $f_f$ the void fraction at failure. Failure is supposed to occur when the void fraction function $f^*$ equals $1/q_1$, thus the damage variable is defined as $D = q_1 f^*$, and the yield function can be rewritten as:
\begin{equation}
\Phi = \left(\frac{\sigma_{eq}}{\sigma_M}\right)^2 + 2 D \cosh \left( \frac{3}{2} q_2 \sigma^*\frac{\sigma_{eq}}{\sigma_M} \right) - (1+D^2).
\label{eq:yield_GTN_2}
\end{equation}

The evolution of the void fraction is governed by both the nucleation of new voids and the growth of existing ones such that the total void fraction growth rate is:
\begin{equation}
\dot{f} = \dot{f}_{nucleation} + \dot{f}_{growth},
\label{eq:GTN_damage_evolution}
\end{equation}
where $\dot{f}_{nucleation}$ and $\dot{f}_{growth}$ represent the nucleation rate and the growth rate of voids, respectively.

Chu and Needleman have suggested that the nucleation rate follow a normal distribution~\cite{CN80}:
\begin{equation}
\dot{f}_{nucleation} = \frac{\mathrm{F}_N}{s_N \sqrt{2\pi}}\exp\left( -\frac{1}{2} \frac{\varepsilon_M^p - \varepsilon_N}{s_N}\right),
\label{eq:GTN_damage_nucleation}
\end{equation}
where $\varepsilon_M^p$ is the undamaged matrix plastic strain, $\varepsilon_N$, $s_N$ are the mean and standard deviation of the distribution of plastic strain respectively, and $\mathrm{F}_N$ is the total void volume fraction that can be nucleated.

The growth rate of voids is comprised of two terms respectively representing that the matrix is plastically incompressible and the mechanism of void softening in shear~\cite{NH08}:
\begin{equation}
\dot{f}_{growth} = (1 - f) \Tr(\bm{\dot{\varepsilon}^p}) + k_{\omega} f \omega(\bm{\sigma})\frac{\bm{\sigma_d}:\bm{\dot{\varepsilon}^p}}{\sigma_{eq}},
\label{eq:GTN_damage_growth}
\end{equation}
where $\bm{\dot{\varepsilon}^p}$ is the macroscopic strain rate, and $\omega(\bm{\sigma})$ a non-dimensional metric defined as:
\begin{equation}
\omega(\bm{\sigma}) = 1 - \left( \frac{27\det(\bm{\sigma_d})}{2\sigma_{eq}^3} \right)^2
\label{eq:omega}
\end{equation}
which lies between $0$ and $1$, with $\omega(\bm{\sigma}) = 0$ for all axisymmetric cases.

Owing to the use of a non-linear yield function, the plasticity algorithm proposed by Leroch \emph{et al.}~\cite{LVEVRG16} needs to be adapted to the GTN model. The new algorithm developed to compute the plastic strain and the deviatoric part of the stress tensor is as follows (Algorithm \ref{alg2}).
At a given time step $n+1$, as it is unknown if yielding has occurred, it is supposed, as a trial, that the total strain increment $\Delta \bm{\varepsilon}$ is purely elastic. The corresponding stress is therefore:
\begin{equation}
\bm{\sigma}_{trial} = \bm{\sigma}_{n} + \bm{C}\Delta\bm{\varepsilon},
\label{eq:sigma_trial_exact}
\end{equation}
where $\bm{C}$ is the stiffness matrix. Since the strain increment is small, the trial von Mises equivalent stress and the trial hydrostatic pressure can respectively be approximated as:
\begin{align}
\sigma_{trial}^{eq} &\approx \sigma_{n}^{eq} \left( 1 + 3G(1-D) \frac{\bm{\sigma}_{n}^d:\Delta\bm{\varepsilon}^d}{{\bm{\sigma}_{n}^{eq}}^2}\right)
\label{eq:sigma_trial_eq_approx}\\
\sigma_{trial}^m &= \sigma_{n}^{m} +  \frac{K}{3} \Tr(\Delta \varepsilon_p),
\label{eq:sigma_trial_m}
\end{align}
where $G$ and $K$ are respectively the shear and the bulk moduli of the undamaged material.

\begin{algorithm} 
\small
\caption{Gurson-Tvergaard-Needleman damage scheme based on the algorithm proposed by Leroch \emph{et al.}\cite{LVEVRG16}.} 
\label{alg2} 
\begin{algorithmic}[1] 
  \STATE $\sigma_{trial}^m = \frac{\rho_0{c_0}^2(\eta -1)[\eta - \frac{\Gamma_0}{2}(\eta - 1)]}{[\eta - S_{\alpha}(\eta-1)]^2} + \Gamma_0e$ \COMMENT{purely elastic pressure update}
  \STATE $\bm{\sigma}_{trial}^d = \bm{\sigma}_{n}^d + 2G(1-D)\Delta\bm{\varepsilon}^d$\COMMENT{purely elastic stress deviator update}
  \STATE $\sigma_{trial}^{eq} = \sqrt{\frac{3}{2} \bm{\sigma}_{trial}^d:\bm{\sigma}_{trial}^d}$\COMMENT{equivalent von Mises trial stress}
  \STATE $\sigma^* = \sigma_{trial}^m / \sigma_{trial}^{eq}$\COMMENT{stress triaxiality}
  \STATE $\sigma_M = \sigma_f(\varepsilon_{M,n}^p)$ \COMMENT{update undamaged matrix flow stress}
  \STATE $\Phi_{trial} = \left(\frac{{\sigma_{trial}^{eq}}}{\sigma_M}\right)^2 + 2 q_1 f^* \cosh \left( \frac{3}{2} q_2 \sigma^*\frac{\sigma_{trial}^{eq}}{\sigma_M} \right) - (1+q_1^2 {f^*}^2)$\COMMENT{trial yield function}
  \IF[yielding did not occur, purely elastic step]{$\Phi_{trial} < 0$}
      \STATE $\bm{\sigma}_{n+1}^d = \bm{\sigma}_{trial}^d$ \COMMENT{keep trial deviatoric stress}
      \STATE $\sigma_{n+1}^m = \sigma_{trial}^m$\COMMENT{keep trial pressure}
  \ELSE[yielding has occurred]
      \STATE $x \leftarrow min(1, \sigma_{trial}^{eq}/\sigma_M)$ \COMMENT{guess that $x = \sigma_{y}/\sigma_M$ is close to $\sigma_{trial}^{eq}/\sigma_M$, but not superior to 1}
      \WHILE[solving yield function using Newton-Raphson]{$|\Delta x|>error$}
          \STATE $\Phi = x^2 + 2 q_1 f^* \cosh \left( \frac{3}{2} q_2 \sigma^*x \right) - (1+q_1^2 {f^*}^2)$
          \STATE $\Phi' = 2x + 3 q_1 q_2 f^* \sigma^* \sinh \left( \frac{3}{2} q_2 \sigma^*x \right)$
          \STATE $\Delta x = -\Phi / \Phi'$
          \STATE $x \leftarrow x + \Delta x$
      \ENDWHILE
      \STATE $\sigma_{y} = x \sigma_M^p$ \COMMENT{update yield stress}
      \STATE $\Delta \varepsilon_{M}^p \leftarrow \frac{\sigma_{trial}^{eq} - \sigma_{y}}{3G(1-D)(1-f)} \left(x + \frac{3}{2}q_1 q_2 f \sigma^* \sinh \left( \frac{3}{2} q_2 \sigma^*x \right)\right)$ \COMMENT{set the undamaged matrix plastic strain increment}
      \WHILE[solving the undamaged matrix plastic strain increment using Newton-Raphson]{$|\delta \Delta \varepsilon_{M}^p|>error$}
          \STATE $\sigma_M = \sigma_f(\varepsilon_{M,n}^p + \Delta \varepsilon_{M}^p)$ \COMMENT{update undamaged matrix flow stress}
          \STATE $g = \Delta \varepsilon_{M}^p - \frac{\sigma_{trial}^{eq} - \sigma_{y}}{3G(1-D)(1-f)} \left(x + \frac{3}{2}q_1 q_2 f \sigma^* \sinh \left( \frac{3}{2} q_2 \sigma^*x \right)\right)$
          \STATE $g' = 1 + x \frac{d\sigma_f}{d\varepsilon_{M}^p}\frac{\left(x + \frac{3}{2}q_1 q_2 f \sigma^* \sinh \left( \frac{3}{2} q_2 \sigma^*x \right)\right)}{3G(1-D)(1-f)} $
          \STATE $\delta \Delta \varepsilon_{M}^p = -g / g'$
          \STATE $\Delta \varepsilon_{M}^p \leftarrow \Delta \varepsilon_{M}^p + \delta \Delta \varepsilon_{M}^p$ \COMMENT{update the undamaged matrix plastic strain increment}
      \ENDWHILE
      \STATE $\varepsilon_{M,n+1}^p = \varepsilon_{M,n}^p + \Delta \varepsilon_{M}^p$\COMMENT{update the undamaged matrix plastic strain}
      \STATE $\sigma_M = \sigma_f(\varepsilon_{M,n+1}^p)$ \COMMENT{update undamaged matrix flow stress}
      \STATE $\sigma_{y} = x \sigma_M$ \COMMENT{update yield stress}
      \STATE $\bm{\sigma}_{n+1}^d = \frac{\sigma_{y}}{\sigma_{trial}^{eq}} \bm{\sigma}_{trial}^d$ \COMMENT{scale deviatoric stress back to yield surface}
      \STATE $\sigma_{n+1}^m = \sigma_{trial}^{m} - 3G(1-D)\sigma^*\Delta \varepsilon_{M}^p$ \COMMENT{scale pressure back to yield surface}
      \STATE $\alpha = \frac{3}{2}\frac{\sigma_M}{\sigma_y} q_1 q_2 f \sinh \left( \frac{3}{2} q_2 \sigma^*x \right)$\COMMENT{ratio $\Tr(\Delta \bm{\varepsilon^p}) / \Delta \varepsilon_p$}
      \STATE $f \leftarrow f + \delta t \frac{\mathrm{F}_N}{s_N \sqrt{2\pi}}\exp\left( -\frac{1}{2} \frac{\varepsilon_{M,n+1}^p - \varepsilon_N}{s_N}\right) + \frac{1 - f}{\alpha \sigma^* + 1} \frac{\sigma_M}{\sigma_y} \left( \alpha(1-f)+ fk_{\omega}\omega(\bm{\sigma})\right) \Delta \varepsilon_{M}^p$ \COMMENT{update the void volume fraction} 
  \ENDIF
\end{algorithmic}
\end{algorithm}

If yield does not occur, $\bm{\sigma}_{trial}$ is exactly the stress tensor at the time step $n+1$, and there is no plastic strain increment. If yielding occurs, however, the yield point is exceeded and the material has deformed plastically. In order to calculate the increment of plastic strain in the undamaged matrix, $\Delta \varepsilon_{M}^p$, first, it is supposed that the void fraction $f$ and the stress triaxiality $\sigma^*$ are constant. Under this assumption, the yield function is only function of the ratio $x = \sigma_{n+1}^{eq}/\sigma_M^p$ and the value of $x$ at which yield occurs is solved numerically using the Newton-Raphson method. Second, since $\Delta \bm{\varepsilon}=\Delta \bm{\varepsilon}_{el}+\Delta \bm{\varepsilon}_p$. The stress tensor at the time step $n+1$ is:
\begin{equation}
\bm{\sigma}_{n+1} = \bm{\sigma}_{n} + \bm{C}\Delta\bm{\varepsilon}_{el},
\label{eq:sigma_n_exact}
\end{equation}
and its equivalent von Mises stress and hydrostatic pressure can be approximated as:
\begin{align}
\sigma_{n+1}^{eq}  = \sigma_y &\approx \sigma_{n}^{eq} \left( 1 + 3G(1-D) \frac{\bm{\sigma}_{n}^d:\Delta\bm{\varepsilon}_{el}^d}{{\bm{\sigma}_{n}^{eq}}^2}\right)
\label{eq:sigma_n_approx}\\
\sigma_{n+1}^m &= \sigma_{n}^{m} +  \frac{K}{3} \Tr(\Delta \varepsilon_p),
\label{eq:sigma_n_m}
\end{align}
where $\sigma_y = x \sigma_M^p$ is the yield stress. By taking the difference between \ref{eq:sigma_trial_eq_approx} and \ref{eq:sigma_n_approx} the equivalent plastic strain increment is obtained as:
\begin{equation}
\Delta \varepsilon_p^{eq} = \frac{\sigma_{trial}^{eq} - \sigma_{y}}{3G(1-D)} = \frac{\sigma_{trial}^{eq} - x\sigma_f(\varepsilon_M^p+\Delta \varepsilon_{M}^p, \dot{\varepsilon}_p, T)}{3G(1-D)}.
\label{eq:delta_epsilon_p_eq}
\end{equation}
Similarly the difference between \ref{eq:sigma_trial_m} and \ref{eq:sigma_n_m} yields
\begin{equation}
\Tr(\Delta \varepsilon_p) = 3\frac{\sigma_{trial}^{m} - \sigma_{n+1}^m}{K},
\label{eq:trace_delta_epsilon_p}
\end{equation}
which is linked to $\Delta \varepsilon_p^{eq}$ using the flow rule
\begin{equation}
\Delta \bm{\varepsilon}_p = \Delta \lambda \frac{\partial \Phi}{\partial \bm{\sigma}} = \Delta \lambda \left[ \frac{3\bm{\sigma}_{n+1}^d}{\sigma_M^2} + \frac{q_1q_2f}{\sigma_M}\sinh\left( \frac{3}{2} q_2 \sigma^*\frac{\sigma_{n+1}^{eq}}{\sigma_M} \right) \bm{I} \right],
\label{eq:flow_rule}
\end{equation}
and gives:
\begin{equation}
Tr(\Delta \bm{\varepsilon}_p) = \alpha \Delta \varepsilon_p^{eq} \quad\textrm{with } \alpha = \frac{3}{2}\frac{\sigma_M}{\sigma_{n+1}^{eq}} q_1 q_2 f \sinh \left( \frac{3}{2} q_2 \sigma^*\frac{\sigma_{n+1}^{eq}}{\sigma_M} \right).
\label{eq:ratio_trace_eq_plastic_strain_increment}
\end{equation}

In order to be able to solve Equation \ref{eq:delta_epsilon_p_eq}, it is assumed that the rate of equivalent plastic work in the matrix material equals the macroscopic rate of plastic work such that~\cite{T89}:
\begin{equation}
\begin{split}
(1-f) \sigma_M \Delta\varepsilon_M^p &= \bm{\sigma}_{n+1}:\Delta\bm{\varepsilon}_p \\
&= \sigma_{n+1}^{eq}(\Delta \varepsilon_p^{eq} + \sigma^* \Tr(\Delta \varepsilon_p)).
\end{split}
\end{equation} 
Thus, solving for $\Delta\varepsilon_M^p$ and plugging it into Equation \ref{eq:delta_epsilon_p_eq} yields
\begin{equation}
\Delta \varepsilon_{M}^p = \frac{\sigma_{trial}^{eq} - x\sigma_f(\varepsilon_M^p + \Delta \varepsilon_{M}^p, \dot{\varepsilon}_p, T)}{3G(1-D)(1-f)} \left(x + \frac{3}{2}q_1 q_2 f \sigma^* \sinh \left( \frac{3}{2} q_2 \sigma^*x \right)\right).
\end{equation}
Finally, this non-linear equation is numerically solved for $\Delta \varepsilon_{M}^p$ also using the Newton-Raphson method.

The final hydrostatic pressure $\sigma_{n+1}^m$ cannot be calculated directly from Equation \ref{eq:sigma_n_m}, since, owing to the non-linear relationship between pressure and deformation, the bulk modulus $K$ is not explicitly known. Locally, a linear approximation of $K$ is obtained by taking the first order approximation of the ratio between Equations \ref{eq:sigma_n_m} and \ref{eq:sigma_n_approx} and recalling that triaxiality is supposed constant:
\begin{equation}
\sigma_{n}^* = \sigma_{n+1}^* \approx \sigma_{n}^* \left[ 1+\left(\frac{3G(1-D)}{\sigma_{trial}^{eq}} - \frac{\alpha K}{3 \sigma_{trial}^m} \right)\Delta \varepsilon_p^{eq} \right] \Leftrightarrow K = \frac{9G(1-D)}{\alpha} \sigma^*.
\end{equation}
Therefore, the new hydrostatic pressure is:
\begin{equation}
\sigma_{n+1}^m = \sigma_{trial}^m - 3G(1-D)\sigma^*\Delta \varepsilon_p^{eq}.
\end{equation}

\section{Case study: simulation of a series of tensile tests of smooth and notched specimens made out of three different Weldox steel alloys, and comparison with FEM results}

\begin{figure}
\begin{center}
\subfloat[]{\includegraphics[height=4cm]{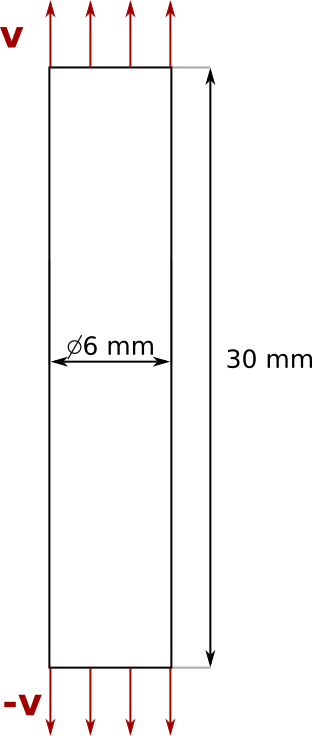}} \hspace{1cm}
\subfloat[]{\includegraphics[height=4cm]{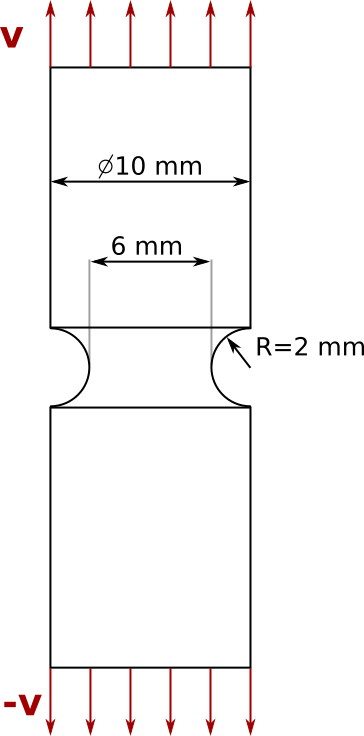}} \hspace{1cm}
\subfloat[]{\includegraphics[height=4cm]{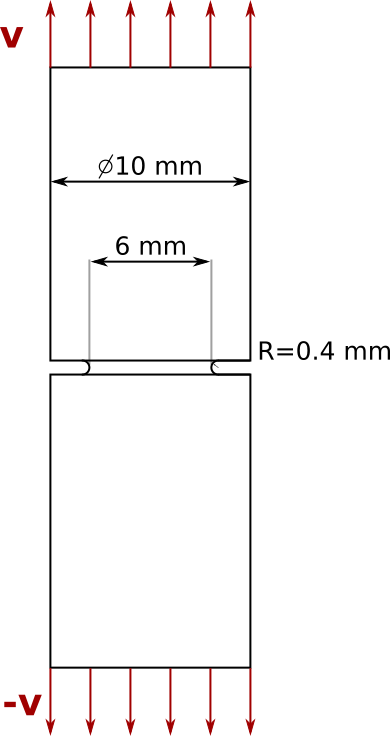}}
\caption{Geometry of (a) the smooth and (b) and (c) the two pre-notched cylindrical specimens simulated tensile specimens~\cite{DBHL06}.}
\label{fig:geometries}
\end{center}
\end{figure}

In order to have confidence in the capabilities of the new damage implementation presented here, it was used to simulate a series of tensile tests of three different specimens each made out of three different Weldox steel alloys. The three different specimen geometries considered are: one smooth and two pre-notched cylindrical specimens with a notch radius of 2~mm and 0.4~mm, respectively (see Fig. \ref{fig:geometries}). Three different Weldox steel alloys are considered: W460E, W700E, and W900E. This case study was chosen because experimental results and material parameters for the three different damage models implemented have been published in the literature~\cite{DBHL06,DBHLL04,OS12}.

For each combination of geometry and alloy, simulations are performed using the CL, the JC, and the GTN damage models as detailed in the previous sections. The results of the TLSPH simulations are then compared with not only the original experimental data published by Dey and coworkers~\cite{DBHL06,DBHLL04}, but also with the results of the same simulations performed using FEM, taken here as a reference. However, the FEM simulations are performed only using the JC damage model as it is the only one of the three damage models to be implemented in the FEM software used here, \emph{i.e} Abaqus Explicit. All the necessary material parameters are taken from the literature. The elasto-plastic parameters as well as the parameters for the CL and JC criteria are those proposed by Dey and coworkers~\cite{DBHLL04,DBHL06} (see Tables \ref{table:mat_param}, \ref{table:mat_const_flow_stress} and \ref{table:mat_const_damage}), while the parameters for the GTN model are taken from {\O}ien and Sch{\o}nberg~\cite{OS12} (see Table \ref{table:mat_const_damage}).

\begin{table}
\begin{center}
\begin{tabular}{cccccc}
\hline
$\rho_0$ [m/s] & $E$ (GPa) & $\nu$ & $c_0$ [m/s] & $S_{\alpha}$ & $\Gamma_0$ \\
\hline
7750 & 211 & 0.33 & 5166 & 1.5 & 0\\
\hline
\end{tabular}
\caption{Material parameters for Weldox steels. $\rho_0$ is the reference bulk density, $E$ the Young modulus, $\nu$ the Poisson's ratio, $c_0 = \sqrt{E/(2(1-2\nu)\rho_0)}$ is the bulk speed of sound, $\Gamma_0$ the  Gr\"uneisen Gamma in the reference state.}
\label{table:mat_param}
\end{center}
\end{table}

\begin{table}
\small
\begin{center}
\begin{tabular}{llll}
\hline
Material & Yield stress & \multicolumn{2}{c}{Strain hardening} \\
 & A (MPa) & B (MPa) & $n$\\
\hline
Weldox 460E & 499 & 382& 0.458\\
Weldox 700E & 859& 329& 0.579\\
Weldox 900E & 992& 364& 0.568\\
\hline
\end{tabular}
\caption{Material constants for the Johnson-Cook constitutive model.\cite{DBHL06}}
\label{table:mat_const_flow_stress}
\end{center}
\end{table}

\begin{table}
\small
\begin{center}
\begin{tabular}{lllllllllll}
\hline
Material & Cockroft-Latham & \multicolumn{5}{c}{Johnson-Cook} & \multicolumn{4}{c}{GTN}\\
 & $W_{cr}$ (MPa) & $D_1$ & $D_2$ & $D_3$ & $D_4$ & $D_5$ & $f_0$ & $f_{cr}$ & $f_F$ & $k_{\omega}$\\
\hline
Weldox 460E & 1219 & 0.636 & 1.936 & -2.969 & -0.0140 & 1.014 & 0.0001 & 0.01 & 0.2 & 5.5\\
Weldox 700E & 1424 & 0.361 & 4.768 & -5.107 & -0.0013 & 1.333 & 0.005 & 0.25 & 0.3 & 3.5\\
Weldox 900E & 1510 & 0.294 & 5.149 & -5.583 & 0.0023 & 0.951 & 0.005 & 0.18 & 0.2 & 3.5\\
\hline
\end{tabular}
\caption{Material constants for three damage and failure models used here. The parameters for the Cockroft-Latham and Johnson-Cook failure criteria were proposed by Dey \emph{et al.}~\cite{DBHL06}, while that for the GTN model were proposed by {\O}ien and Sch{\o}nberg~\cite{OS12}. }
\label{table:mat_const_damage}
\end{center}
\end{table}

In all simulations, the full gauge length of the smooth specimens is simulated, while, to minimize computational times, for the pre-notched specimens, only a portion of respectively 7~mm and 20~mm in height around the notch is considered. In SPH, all domains are discretized using a simple cubic array of particles which constant spacing is of 0.15~mm and 0.3~mm for the smooth and $R=2~mm$ notched specimen, respectively, while for the $R=0.4~mm$ notched specimen it varies from 0.8~mm for furthest point from the notch, to 0.1~mm in the notch. On the last top and bottom rows of particles, a velocity $\pm v$ of the form $ v_{max}(1-e^{-t})$ is applied in the direction colinear to the each specimen's axis (Fig. \ref{fig:geometries}). As no strain rate effect is considered here, to speed-up simulations, a high maximum velocity is chosen: $v_{max} = 10~mm/ms$. Such a high velocity was chosen in order to decrease the total required time to be simulated and has no impact on the results as no strain rate effect was taken into account.

\subsection{Results}

\begin{figure}[H]
\begin{center}
\subfloat[]{\includegraphics[width=13cm]{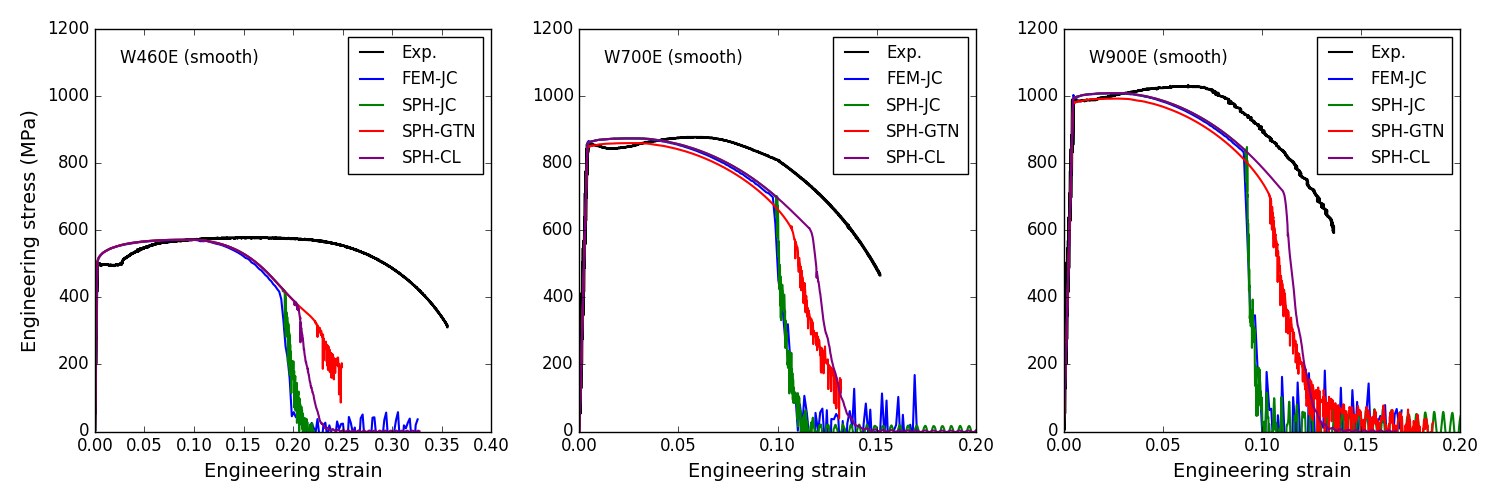}\label{fig:stress-strain_curves_smooth}}\\
\subfloat[]{\includegraphics[width=13cm]{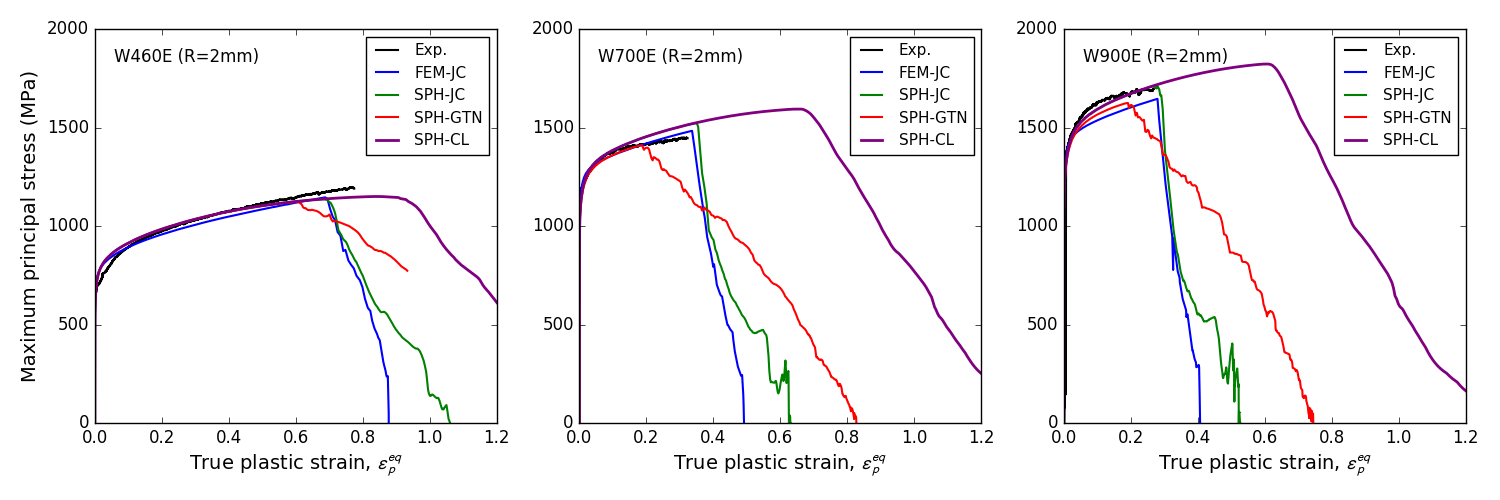}\label{fig:stress-strain_curves_R_2}}\\
\subfloat[]{\includegraphics[width=13cm]{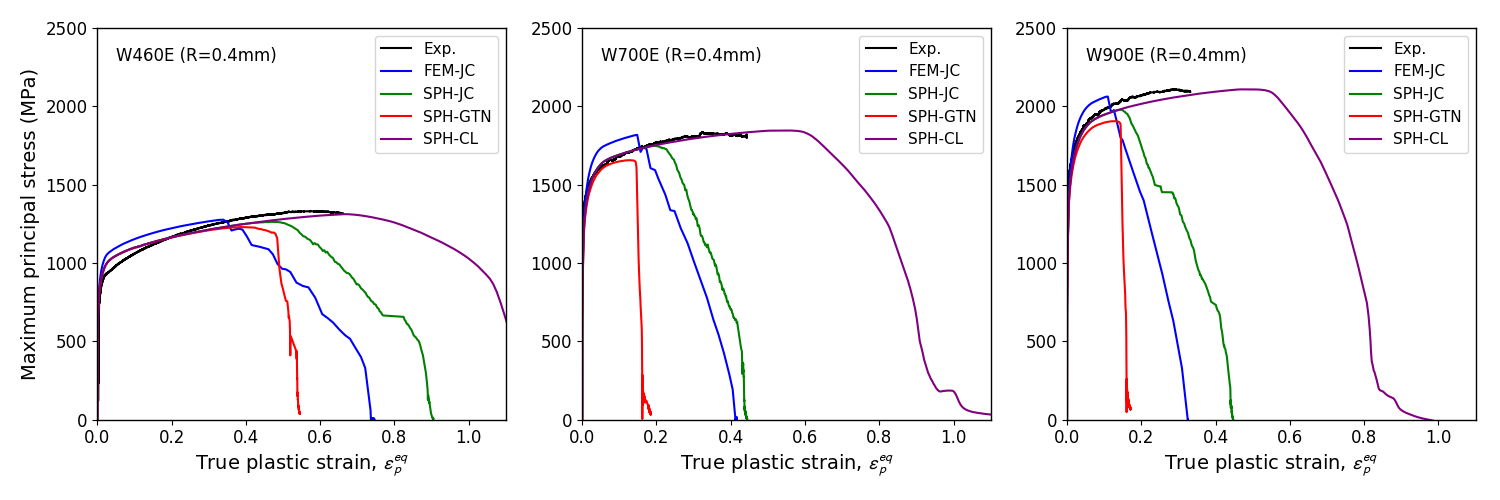}\label{fig:stress-strain_curves_R_0.4}}
\caption{Comparison of the stress-strain curve of the smooth and pre-notched cylindrical specimens for Weldox 460E, Weldox 700E, and Weldox 900E. In black are plotted the experimental results from Dey \emph{et al.}~\cite{DBHL06}, in blue the results of finite element simulations using Johnson-Cook damage model and in green, red, and purple, the results of SPH simulations using Johnson-Cook, Cockroft-Latham and Gurson-Tvergaard-Needleman damage models, respectively.}
\label{fig:stress-strain_curves}
\end{center}
\end{figure}

Fig. \ref{fig:stress-strain_curves} presents the results of the SPH simulations for all combination of specimen geometries and alloys tested plotted alongside both the experimental data and FEM results. More specifically, Fig. \ref{fig:stress-strain_curves_smooth} shows the evolution of the engineering stress as a function of the engineering strain for the smooth specimens, while Fig. \ref{fig:stress-strain_curves_R_2} and \ref{fig:stress-strain_curves_R_0.4} show the variation of principal stress vs. true plastic strain averaged over the cross-section at the middle of the notch of both pre-notched specimens. 

First, it can be seen that, the flow stresses predicted by the SPH simulation is in good agreement overall with the FEM. The agreement is very good in the case of the smooth and $R=2~mm$ notched specimens, while for the $R=0.4~mm$ notched specimens not as good agreement is obtained. In that case, the results of the SPH simulations fit better the experimental data than FEM. In fact, FEM consistently overpredicts the flow stresses for all three alloys. This results is, however, consistent with the results of the FEM simulations performed by Dey and coworkers~\cite{DBHL06,DBHLL04}.
It can be also noticed that for all these simulation cases, the flow stress given by SPH using the GTN model is slightly lower than that obtained with the other damage models. This is due to the fact that GTN uses a different yield function and that presence of voids in the underformed material is considered, \textit{i.e.} $f_0 \neq 0$.

Second, as expected, after damage initiates, a steady decay in stress is observed, and the simulations are numerically stable. The strains at failure predicted by SPH using the JC model are in very good agreement with those predicted using FEM in the case of the smooth and $R=2~mm$ notched specimens. In the case of the $R=0.4~mm$ notched specimens, however, SPH simulations predict strain at failure that are higher than those predicted by FEM, but they are a little closer to experimental results. Such discrepancies can be explained by the lowest flow stress observed in the SPH simulations. The strains at failure predicted by SPH using the CL model, however, are always higher than that using JC. This is consistent with conclusions from Dey and coworkers' work~\cite{DBHL06,DBHLL04}.

Since the results obtained using the Johnson-Cook damage model using SPH and FEM are in good agreement with each other, we have confidence that the damage implementation in SPH is successful.
Also, we could see that, none of the models are quantitatively able to predict the experimental results. This was expected as all the parameters used were taken directly from the literature and efforts were not made to optimise them from the point of view of agreement with experiment. Of course, this is an important aspect of using SPH to simulate real experimental data but that is not the focus of this contribution, which is on presenting this new implementation of damage and failure in SPH.

\begin{figure}
\hspace{-2cm}
\begin{tikzpicture}
\node at (-7,0) {\includegraphics[width=5.5cm]{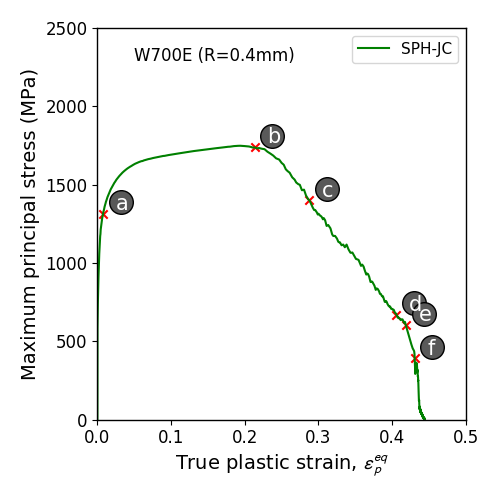}};
\node at (-3,3) {\includegraphics[width=3cm]{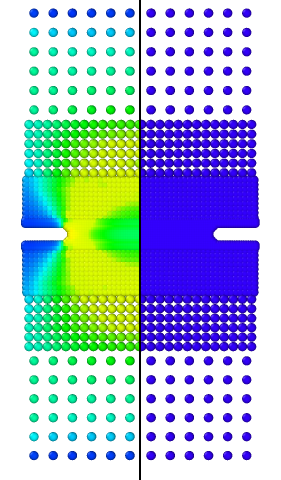}};
\node at (-3.75,0.5) {\tiny \textbf{von Mises}};
\node at (-3.75,0.3) {\tiny \textbf{stress}};
\node at (-2.4,0.5) {\tiny \textbf{Damage}};
\node at (0,3) {\includegraphics[width=3cm]{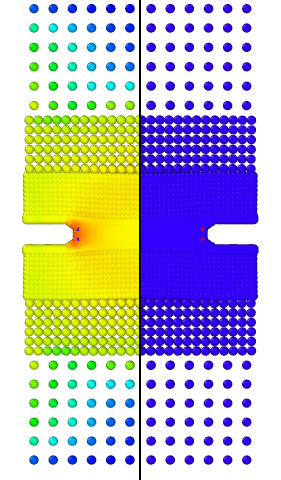}};
\node at (-0.75,0.5) {\tiny \textbf{von Mises}};
\node at (-0.75,0.3) {\tiny \textbf{stress}};
\node at (0.6,0.5) {\tiny \textbf{Damage}};
\node at (3,3) {\includegraphics[width=3cm]{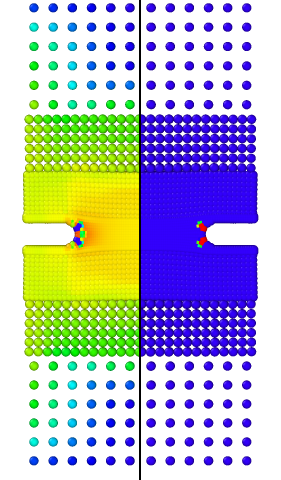}};
\node at (2.25,0.5) {\tiny \textbf{von Mises}};
\node at (2.25,0.3) {\tiny \textbf{stress}};
\node at (3.6,0.5) {\tiny \textbf{Damage}};
\node at (-3,-3) {\includegraphics[width=3cm]{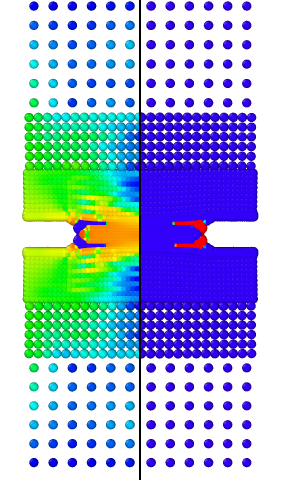}};
\node at (-3.75,-5.5) {\tiny \textbf{von Mises}};
\node at (-3.75,-5.7) {\tiny \textbf{stress}};
\node at (-2.4,-5.5) {\tiny \textbf{Damage}};
\node at (0,-3) {\includegraphics[width=3cm]{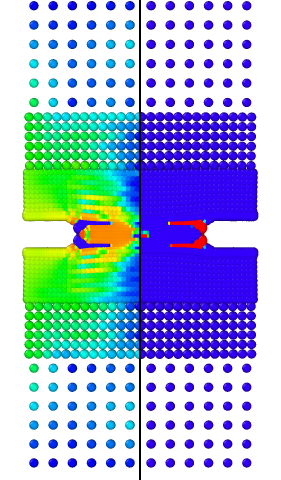}};
\node at (-.75,-5.5) {\tiny \textbf{von Mises}};
\node at (-.75,-5.7) {\tiny \textbf{stress}};
\node at (0.6,-5.5) {\tiny \textbf{Damage}};
\node at (3,-3) {\includegraphics[width=3cm]{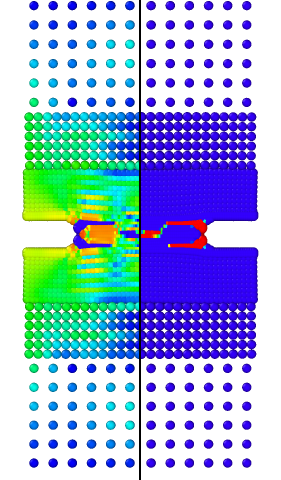}};
\node at (2.25,-5.5) {\tiny \textbf{von Mises}};
\node at (2.25,-5.7) {\tiny \textbf{stress}};
\node at (3.6,-5.5) {\tiny \textbf{Damage}};
\node at (5.5,0) {\includegraphics[width=2cm]{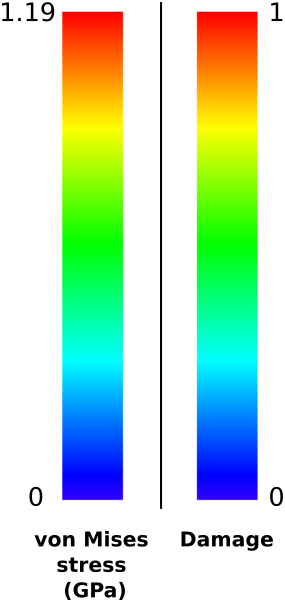}};
\node [shape=circle,draw,inner sep=1.5pt,fill=Grey] at (-3,5) {\footnotesize\textcolor{white}{\textbf{a}}};
\node [shape=circle,draw,inner sep=1.5pt,fill=Grey] at (0,5) {\footnotesize\textcolor{white}{\textbf{b}}};
\node [shape=circle,draw,inner sep=1.5pt,fill=Grey] at (3,5) {\footnotesize\textcolor{white}{\textbf{c}}};
\node [shape=circle,draw,inner sep=1.5pt,fill=Grey] at (-3,-1) {\footnotesize\textcolor{white}{\textbf{d}}};
\node [shape=circle,draw,inner sep=1.5pt,fill=Grey] at (0,-1) {\footnotesize\textcolor{white}{\textbf{e}}};
\node [shape=circle,draw,inner sep=1.5pt,fill=Grey] at (3,-1) {\footnotesize\textcolor{white}{\textbf{f}}};
\end{tikzpicture}
\caption{Details of the evolution of both the equivalent von Mises stress and damage distributions obtained the new TLSPH simulations of a W700E notched cylindrical specimen which notch radius is of $R=0.4~mm$ using the Johnson-Cook damage criterion. The snapshots show orthogonal cross-sectional views of the simulated specimen along its axis of symmetry and are divided into two parts: on the left half the particles' colors are function of the equivalent stress, while on the right half they are function of the damage level. These snapshots were created using OVITO~\cite{S10}. Please note that the radius of the particles as displayed is constant and independent of their respective volume.}
\label{fig:damage_evolution}
\end{figure}
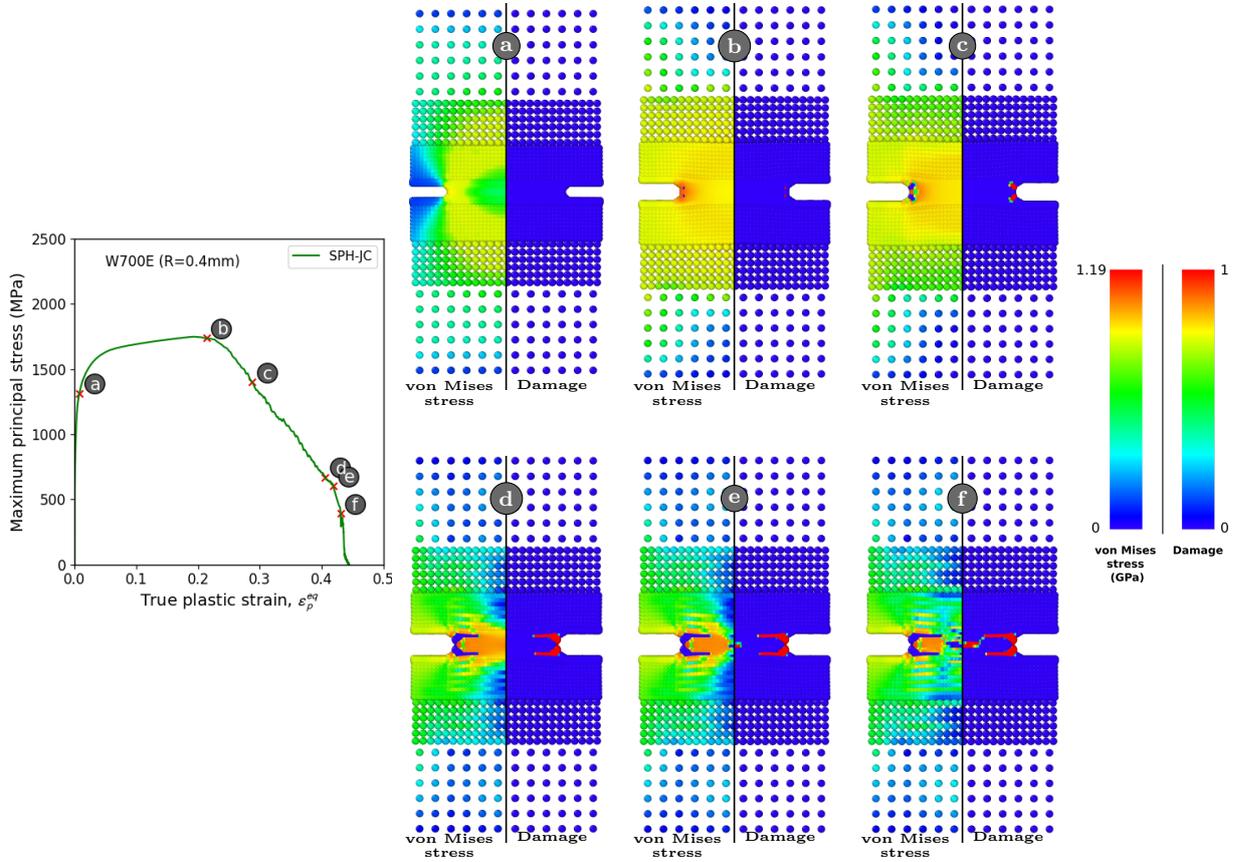

A typical example of the result of a TLSPH simulation is given by Fig. \ref{fig:damage_evolution}. It shows, alongside the stress-strain curve, the evolution of distribution of both the equivalent von Mises stress and damage accumulated in a W700E notched cylindrical specimen which notch radius is of $R=0.4~mm$ using the JC damage criterion. At the onset of plasticity (snapshot (a)), it can be seen that, as expected, the stress is concentrated around the notch tip. Thus, it is natural to see that when the strain at failure is reached (snapshots (b)) that cracks initiate from this same point. One can also clearly distinguish from snapshot (c) that the two forming cracks are oriented at $\pm 45^\circ$ with the horizontal, \textit{i.e.} in the direction of the principal stresses, and that their presence lead to a decrease of maximum principal stress. The stress relaxation in the outer parts of the specimen that follow cause a change in the crack growth direction (snapshot (d)) as they start propagating horizontally. This further growth lead to stress at the center of the specimen along a horizontal plane. There, another crack initiates (snapshot (e)) and quickly propagates outwards to connect with the upper crack and break the specimen into two (snapshot (f)).

\section{Discussion}

For the simulations of mechanical problems involving large deformations, particle based methods present the advantage over Finite Elements Modeling to not require re-meshing which is very CPU intensive, and is not always guaranteed to be successful. When using particle based methods for such applications, however, one needs to be aware of the risk of numerical failure that would occur when the distance between two neighbouring particles becomes eventually higher than the the kernel cutoff radius. This can lead to failure strains and energy releases that are not physical. This is a reason why TLSPH is more favourable than classical SPH for simulations where their is a probability of failure. In fact, in TLSPH, if the reference configuration is never updated, two neighbouring particles would remain as such whatever their separation distance in the deformed state, thus preventing numerical failure from occurring.

Using TLSPH with the ``pseudo-spring'' method for the simulation of ductile materials generates numerical instabilities as soon as damage initiates. This triggered the need for a new way of taking damage into account in the SPH approximation.
The resulting new implementation was tested on a series of tensile tests for different damage criteria. The results of these tests show that, due to the good agreement with FEM and experimental results, the new TLSPH approximation can be trusted. They also show that, depending on the type of specimen and elasto-plastic parameters used, some damage criteria give better results than others. The Johnson-Cook, Cockcroft-Latham, and Gurson-Tvergaard-Needleman damage criteria have been implemented as they are one of the most popular criteria used in academia and industry. These changes have all been implemented in the user package SMD of LAMMPS for the community to test and use. The code is available in the first author's Github account: \url{https://github.com/adevaucorbeil/lammps/} .


\section{Conclusion}

A new method for damage and failure was implemented in the LAMMPS user package Smooth Mach Dynamics for the simulation of ductile solids using Total-Lagrangian SPH. Three popular damage criteria that are Cockroft-Latham, Johnson-Cook, and Gurson-Tvergaard-Needleman have also been implemented in this package. This new method was then tested on the study of tensile tests of smooth and notched specimens made out of three different Weldox steel alloys. The simulation results obtained using TLSPH were then compared to both experimental and FEM results. These results shows that TLSPH can be used successfully to simulate the mechanical response of materials experiencing damage and failure without relying on or exploiting numerical failure. The simulations presented here serve as validation of the implementation and are the first necessary step before being able to use this model for the simulation of more complicated phenomena such as machining, wear, and impacts of ductile materials.

\section*{Declaration of Interest}
The authors declare that they have no known competing financial interests or personal relationships that could have appeared to influence the work reported in this paper.

\section*{Acknowledgments}

The authors gratefully acknowledge the financial support of the Australian Research Council (ARC) Training Centre in Alloy Innovation for Mining Efficiency (IC160100036). Fruitful discussions with Prof. Matthew Barnett (Deakin University and Director of the ARC Training Centre) are acknowledged.

\section*{References}
\bibliographystyle{elsarticle-num}
\bibliography{literature}

\end{document}